\documentclass[letterpaper]{article} 
\usepackage{aaai25}  
\usepackage{times}  
\usepackage{helvet}  
\usepackage{courier}  
\usepackage[hyphens]{url}  
\usepackage{graphicx} 
\usepackage{natbib}  
\usepackage{caption} 
\usepackage{algorithm}
\usepackage{algorithmic}

\usepackage{booktabs}
\usepackage{tabularx}

\usepackage{booktabs}
\usepackage{multirow}
\usepackage{color}
\usepackage{xcolor}
\usepackage{colortbl}

\usepackage{comment}

\usepackage{siunitx}

\usepackage{marvosym}

\usepackage{xcolor}

\usepackage{newfloat}
\usepackage{listings}
\DeclareCaptionStyle{ruled}{labelfont=normalfont,labelsep=colon,strut=off} 
\floatstyle{ruled}
\newfloat{listing}{tb}{lst}{}
\floatname{listing}{Listing}

\title{\emph{Telegram as a Battlefield:}\\
Kremlin-related Communications during the Russia-Ukraine Conflict}
\author{
    Apaar Bawa\textsuperscript{\rm 1},
    Ugur Kursuncu\textsuperscript{\rm 1}\textsuperscript{\Letter},
    Dilshod Achilov\textsuperscript{\rm 2},
    Valerie L. Shalin\textsuperscript{\rm 3},\\
    Nitin Agarwal\textsuperscript{\rm 4,5},
    Esra Akbas\textsuperscript{\rm 1}
}
\affiliations{

    \textsuperscript{\rm 1}Georgia State University, 
    \textsuperscript{\rm 2}University of Massachusetts Dartmouth, 
    \textsuperscript{\rm 3}Wright State University,\\
    \textsuperscript{\rm 4}COSMOS Research Center, University of Arkansas - Little Rock\\
    \textsuperscript{\rm 5}International Computer Science Institute, University of California, Berkeley\\
    abawa2@student.gsu.edu, ugur@gsu.edu, dachilov@umassd.edu, valerie.shalin@wright.edu, nxagarwal@ualr.edu, eakbas1@gsu.edu
}

\begin{document}

\maketitle

\begin{abstract}
Telegram emerged as a crucial platform for both parties during the conflict between Russia and Ukraine. Per its minimal policies for content moderation, Pro-Kremlin narratives and potential misinformation were spread on Telegram, while anti-Kremlin narratives with related content were also propagated, such as war footage, troop movements, maps of bomb shelters, and air raid warnings. This paper presents a dataset of posts from both pro-Kremlin and anti-Kremlin Telegram channels, collected over a period spanning a year before and a year after the Russian invasion. The dataset comprises $404$ pro-Kremlin channels with $4,109,645$ posts and $114$ anti-Kremlin channels with $1,117,768$ posts. We provide details on the data collection process, processing methods, and dataset characterization. Lastly, we discuss the potential research opportunities this dataset may enable researchers across various disciplines. 
\end{abstract}


\section{Introduction}

Social media has become a primary communication tool for political communication and influence, serving both state-affiliated actors and opposition groups in shaping public discourse and mobilizing support. In the Russian context, the government has leveraged online platforms to promote its policy agenda, manage its international image, and suppress dissenting narratives \citep{Gunitsky2015,Spaiser2017}. These online efforts build upon the Kremlin's long-standing control over traditional media, including state-owned television channels, such as Rossiya 1 and NTV, which have been instrumental in disseminating official narratives \citep{Aro2016}. Prior research has highlighted how these strategies extend to online communications where disinformation campaigns, surveillance measures, and restrictions on independent journalism have been employed to limit political pluralism and suppress pro-democratic discourse \citet{Polyanskaya2003,Aro2016}. This extensive ability to control communications enables the state to exert significant influence over public opinion.

On the other hand, the Russian opposition (i.e., anti-Kremlin) relies on social media platforms with more flexible content moderation policies, such as Telegram \citep{Jurcevic2019}. Despite increased restrictions, digital platforms have amplified the voices of the Kremlin's critics \citep{Glazunova2023}. \citet{Smyth2017} have concluded that while state media's dominant narrative fosters regime support, alternative debates and government critiques on new media form the cognitive basis for opposition. For instance, intelligence on Russian troops’ border movements and military plans was leaked and became trending on social media just before President Vladimir Putin announced the full-scale invasion \citep{Karalis2024}. Because anti-Kremlin narratives can only be spread on such online platforms, it is crucial for researchers to investigate the dynamics and potential strategies of these communications.

\subsection{Russian Invasion of Ukraine Portrayed in Online Communications}

In February 2022, as Russian forces invaded Ukraine, the Russian government swiftly enacted legislation aimed at controlling the narrative around the conflict. This included a "fake news" law that specifically banned the use of the term "war" to describe the invasion, requiring instead the use of the government-approved phrase "special military operation" \citep{Sherstoboeva2024}. Violations of this law led to severe punishment, including imprisonment for up to ten years, the shutdown of numerous news outlets, and the criminal prosecution of thousands of individuals for alleged misinformation. Pro-Kremlin channels often emphasized the legitimacy of the "special military operation," focusing on liberation and national security. Persisting anti-Kremlin channels highlighted war atrocities, humanitarian crises, and resistance efforts \cite{Ramani2023}. The role of social media in this conflict is not merely passive; it actively attempts to shape perceptions and mobilize support or dissent. The rapid spread of information, from verified reports to misinformation, underscores the importance of digital literacy and critical analysis in contemporary conflicts \citep{pierri2023propaganda}. 

\subsection{Telegram as an Essential Propaganda Medium}

Major online platforms such as Facebook and X (formerly Twitter) have become critical sources of information for news, activism, business, and marketing in the last two decades \citep{Kalsnes2018,Kursuncu2021}. Similarly, Telegram, an instant messaging app, has gained substantial popularity, especially in Russia and Eastern Europe, where it serves as a crucial alternative to state-monitored communication channels for sharing news \citep{Khaund2020}. With Russia's invasion of Ukraine, Facebook and Instagram were banned over extremism charges by a Moscow court, and access to Twitter was restricted by the Russian censorship body Roskomnadzor \citep{Glazunova2023}. In that light, Telegram's role as a vital source of information has become increasingly significant. According to a survey by the Levada \citet{Levada2023}, among Russian youth aged $18-24$, TV was found to be the least popular news source, with $41\%$ relying on Telegram channels for news. Additionally, during the initial two months of the conflict, $76.6\%$ of Ukrainians used social media for news, with Telegram being the most favored platform by 65.7\% \citep{OPORA2022}. Both Russian and Ukrainian government officials also actively use Telegram to rally international support, broadcast air raid warnings, and distribute maps of local bomb shelters.

Telegram distinguishes itself from other major online platforms through several unique features: (i) Telegram channels can be public or private, serving as feeds where text, photos, videos, audio, documents, and polls are broadcast to large audiences, including unlimited subscribers. (ii) Unlike other platforms, Telegram does not use algorithms to prioritize or restrict content visibility, giving users more straightforward access to all posted content. (iii) Users can easily disable comments and reactions, making channels a one-directional communication tool for mass information dissemination. These features have positioned Telegram as an effective platform for both anti- and pro-Kremlin propaganda within Russia, leading to its wide adoption for news consumption and dissemination. Over $40\%$ of all Telegram channels are dedicated to news, with approximately $17\%$ directly related to the war and ongoing events in Ukraine \citep{ReRussia2024}. This also makes Telegram a central medium for real-time updates and reliable information, influencing public opinion and information flow. 

The Internet Research Agency (IRA), a Russian organization often associated with Kremlin-linked information operations, has been identified as playing a central role in coordinated online influence campaigns by disseminating polarizing narratives on Western online platforms \citep{alieva2024analyzing, linvill2020troll,starbird2019disinformation}. While much of the IRA's documented activity has focused on platforms such as X/Twitter and Facebook, growing evidence indicates its activity on Telegram as well \cite{ukgov2022trollfactory}. In February 2023, a Telegram channel affiliated with the Wagner Group published a statement from Yevgeny Prigozhin acknowledging his connection to the IRA \cite{krever2023wagner}. This development adds to concerns regarding Telegram’s use as an instrument for coordinated influence operations, besides being an online space for independent communication.

\section{Related Work}

A growing body of research has examined online communications related to the Russian-Ukraine conflict across online platforms including X/Twitter, Reddit and Telegram. Much of this literature focuses on the spread of propaganda, disinformation, and strategic narratives by various state and non-state actors. 

\citet{hanley2023special} conducted a comparative analysis of Western,  Russian, and  Chinese media ecosystems by examining 11,359 news articles and associated social media content published between January and April 15, 2022. Their work highlights distinct thematic patterns in how the conflict was framed in traditional and social media across these regions. In another study \cite{hanley2023happenstance}, the same authors explored the dissemination of Russian state-sponsored narratives to English-speaking audiences by analyzing content from nine English-language Russian media websites and related Reddit communities, including r/Russia and 10 other political subreddits, to understand the most prominent narratives touted by the Kremlin for English-speaking audiences. In a subsequent work, \citet{hanley2024partial} curated 732 Telegram channels hyperlinked by 16 Russian media websites tracked by the U.S. State Department, providing insight into how narratives circulate between formal news sites and Telegram. However, this dataset largely comprises pro-Kremlin sources and is limited to a specific timeframe.  

Beyond Telegram, multilingual social media analyses have been conducted. For instance, \citet{lai2024multilingual} curated a X/Twitter dataset containing approximately 53 million tweets in six languages (English, Japanese, Spanish, French, German, Korean) during the early weeks of the invasion, identifying narratives and misinformation trends across linguistic communities.

Other researchers have focused on building datasets from Telegram to support broader interest areas. The Pushshift Telegram Dataset \citet{baumgartner2020pushshift} is one of the largest publicly available collections, encompassing 317 million messages from 27,801 channels and over 2.2 million users. Though it predates the current conflict, its infrastructure informed subsequent data collection efforts. \citet{gruzd2024shaping}, funded by Canada’s Social Sciences and Humanities Research Council (SSHRC), compiled data from 55 English-language Telegram channels covering the conflict between February 2022 and February 2023, although details of their sampling strategy remain limited.

Telegram continues to be leveraged in research on specific narratives and movements. \citet{herasimenka2022movement} used Telegram data to study leadership strategies in anti-authoritarian protest movements, focusing on Alexei Navalny’s 2017 mobilization in Russia. \citet{kloo2024cross} assembled a dataset of 6.7 million Telegram messages from over 18,000 channels to analyze disinformation narratives, particularly those portraying Ukraine as a “Nazi state.” \citet{ghasiya2023messaging} examined communication patterns on Telegram during the conflict by focusing on selected high-profile channels from both Ukrainian and Russian sides, such as UkraineNow and RT Russian.

\subsection{The Current Study}

This paper presents the first comprehensive Telegram dataset capturing both pro-Kremlin and anti-Kremlin Telegram channels within the context of the Russian political landscape. Our dataset is created from 404 pro-Kremlin and 114 anti-Kremlin Telegram channels, with over four million and one million posts, respectively. These channels have been instrumental in disseminating various narratives and counter-narratives within the ongoing Ukraine conflict \citep{ISD2022,bawa2024adaptive}. A portion of this dataset, covering anti-Kremlin channels from January 2022 to March 2023, has been analyzed to examine the dynamics between offline events and online communications in anti-Kremlin channels over the seven distinct phases of the invasion \citep{bawa2024adaptive}, as outlined by \citet{Murauskaite2023}. In the processes of data collection and curation, we ensure transparency and reproducibility in our approach. We provide details on data characteristics, including post volumes, views, forwarding and user engagement levels, and the use of multimodal content (text, images, videos). We also present n-gram analysis to provide insights into the lexical and topic characteristics prevalent in data, enabling parallel comparisons between the channels \citep{kursuncu2019predictive}. This unique dataset serves as an essential resource for understanding the dynamics of online political discourse in Russia, particularly in how digital platforms are used to influence public opinion and engage with the audience during periods of escalated political concern.

\section{Methods}
This section outlines our methodology for collecting and analyzing data from Russian political Telegram channels during the Russia-Ukraine conflict. We describe the process of selecting and categorizing channels, detail the data collection techniques, and explain how channel labels were validated. In addition, we present an overview of the dataset, including the types of attributes collected, covering the period before and after the invasion. Finally, we provide a content analysis highlighting the differences between Pro-Kremlin and Anti-Kremlin channels.

\begin{figure}[hbt!]
  \centering
  \includegraphics[width=8.5cm]{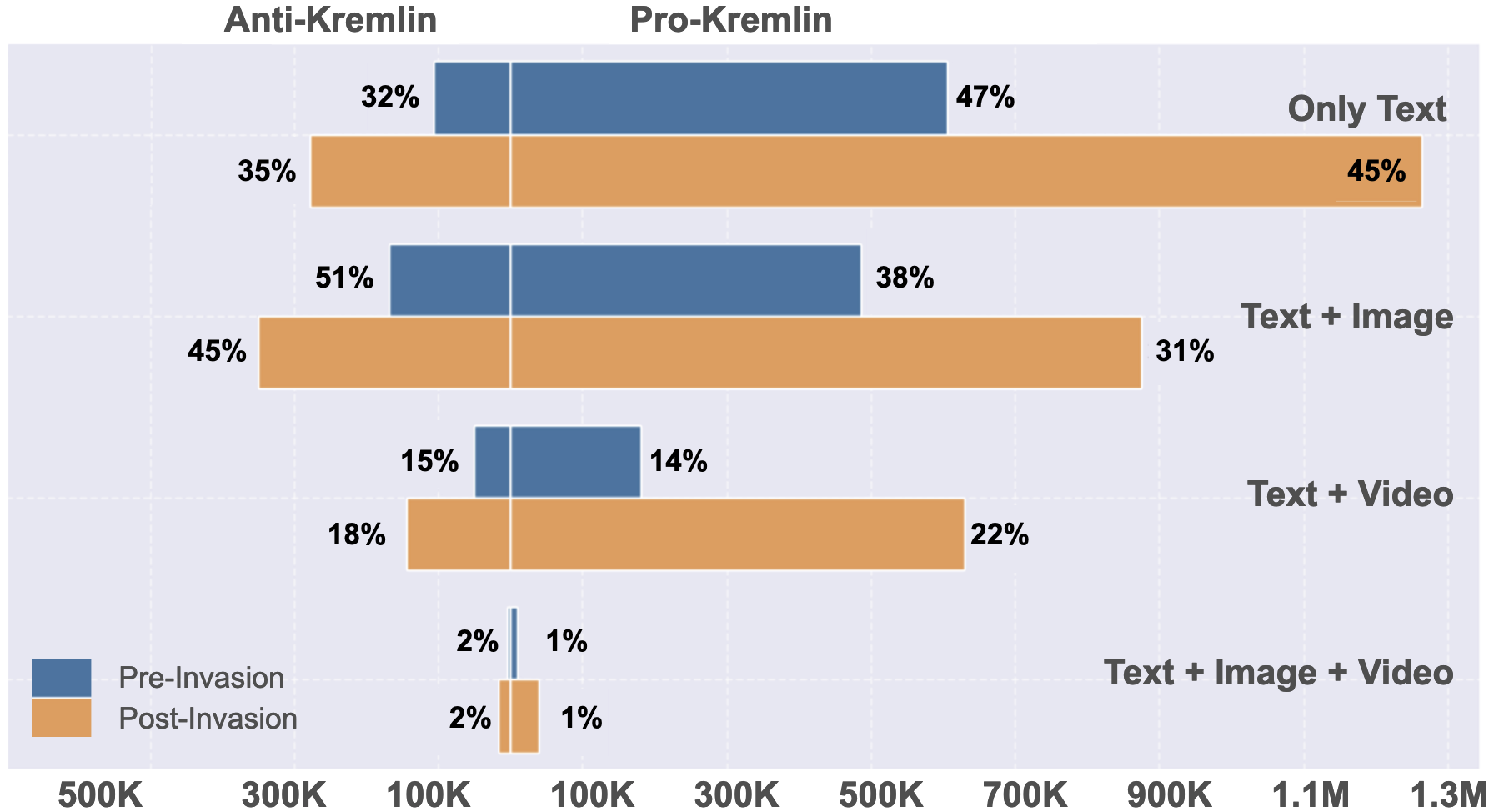}
\caption{Comparison of Anti-Kremlin and Pro-Kremlin post volumes across different modalities (Only Text, Text Image, Text Video, and Text Image Video) during the Pre-Invasion and Post-Invasion phases.}
\label{fig:post-volume-modality}
\end{figure} 

\subsection{Data Collection and Annotation}
\label{sec:data-collection-and-annotation}

We collected Telegram data using the open-source Python library Telethon 
\citep{Telethon2024}, which provides client access to Telegram's public API. This API supports the retrieval of publicly available content such as messages, metadata, and user/channel information from public channels and groups, while excluding private or deleted content. The library adheres Telegram's rate limits, including ~200 username per day. Our data collection was conducted retrospectively in batches, starting in May 2023 and continuing until June 2023.

To identify relevant Russian-language political Telegram channels, we used the TGStat platform \citep{TGStat2024}, which catalogs channels by country, language, and thematic content. We selected channels with at least $10,000$ subscribers to focus on influential accounts likely to contribute to public discourse. 
These channels were manually labeled by a native Russian-speaking coder into four categories: Pro-Kremlin, Anti-Kremlin, Neutral, and Others. Pro-Kremlin channels were identified based on their consistent dissemination and frequent alignment with official Kremlin narratives and support for state policies. In contrast, anti-Kremlin channels were those exhibiting critical stance toward the policies and actions of Russian government actions and leadership. Channels deemed “neutral” predominantly shared news without discernible political bias. The “Others” category included topics unrelated to the focus of our study, such as other countries' politics. 

Label validation was conducted in two stages. First, a non-Russian-speaking coder reviewed the classifications using Telegram's built-in translation tools. Second, a native Russian-speaking political science expert reviewed a randomly selected subset of $100$ channels to assess the label accuracy and reliability. To evaluate the inter-rater agreement, we calculated Cohen's kappa coefficient \citep{Cohen1960}, $\kappa$. While $\kappa$ provides a measure of consistency across coders, it does not confirm the accuracy of labels and should be interpreted accordingly.

\subsection{Data Description}
\label{sec:data-description}

Following the annotation process, we identified $404$ pro-Kremlin and $114$ anti-Kremlin channels. Data collection from these channels spanned $430$ days from December $21$, $2020$, to April $30$, $2023$—roughly more than a year pre-invasion and post-invasion. The dataset includes the following attributes of Telegram posts: (i) main text content of the posts, (ii) accompanying media (e.g., images, videos –represented as placeholders in the dataset), (iii) timestamps (e.g., date and time), (iv) number of view counts per post, (v) post forwarding counts (i.e., how often the post is forwarded), (vi) original or forwarded status (i.e., whether a post is original or forwarded), (vii) forwarding source (i.e., the origin source of the forwarded post, if applicable), (viii) emoji reactions (i.e., type of emoji reactions and their frequencies), and (ix) user replies to the post (i.e., each reply contains similar data as above). This comprehensive dataset contains $4,109,645$ posts from pro-Kremlin channels and $1,117,768$ posts from anti-Kremlin channels. The inclusion of such diverse data attributes allows for an in-depth analysis of communication patterns and user engagement on these channels.	

\begin{table*}[ht]
\centering
\begin{tabular}{lrrrrrr}
\toprule[2pt]
\textbf{Modalities} & \multicolumn{3}{c}{\textbf{Pro-Kremlin}} & \multicolumn{3}{c}{\textbf{Anti-Kremlin}} \\
\cmidrule(r){2-4} \cmidrule(r){5-7}
                    & \textbf{Pre-Invasion} & \textbf{Post-Invasion} & \textbf{\% Change} & \textbf{Pre-Invasion} & \textbf{Post-Invasion} & \textbf{\% Change} \\
\midrule[1pt]
\textbf{Only Text}        & 18,696 & 62,853 & 236.2\% & 25,006 & 101,597 & 306.3\% \\
\textbf{Text + Image}     & 12,106 & 59,333 & 390.1\% & 25,218 & 103,115 & 308.9\% \\
\textbf{Text + Video}     & 16,957 & 86,447 & 409.8\% & 26,355 & 112,429 & 326.6\% \\
\textbf{Text + Image + Video} & 20,529 & 97,394 & 374.4\% & 26,424 & 127,480 & 382.4\% \\
\bottomrule[2pt]
\end{tabular}
\caption{Distribution of the average number of views for different modalities, Pre- and Post-Invasion for both Pro-Kremlin and Anti-Kremlin channels.}
\label{tab:modalities_views}
\end{table*}

\section{Data Records}

We performed a comparative exploratory analysis of both Pro-Kremlin and Anti-Kremlin channels in the context of pre- and post-invasion time frames. We examined the following attributes of the posts: (i) Multimodalities and Views (i.e., presence of various modalities of data and view counts),  (ii) Post Forwarding, (iii) Reactions, and (iv) Replies.

\subsection{Multimodality and Views}
\label{sec:multimodalities-and-views}

Research shows that multimodal content, a critical feature that Telegram users often utilize \citep{MoscowTimes2023}, boosts user engagement \citep{Park2022,Voorveld2018}. Both Pro-Kremlin channels and anti-Kremlin channels demonstrate notable increases in post volumes, from $1,289,173$ pre-invasion to $2,820,472$ post-invasion by $119\%$, and from $330,492$ to $787,276$ by $138\%$, respectively. Figure \ref{fig:post-volume-modality} presents a comparison across pre- and post-invasion post volumes between Pro- and Anti-Kremlin channels, highlighting changes in their multimodal characteristics. The content is various combinations of modalities as follows; (i) text and image, (ii) text and video, (iii) text, image and video, (iv) only text. The relative proportions of each modality type for a given group and period reveal cues on evolving communication strategies. For instance, text-only posts in Pro-Kremlin channels decreased slightly from 47\% to 45\% of total posts despite an absolute increase in count, while posts combining text and video increased significantly. In contrast, Anti-Kremlin channels experienced a modest 3\% rise in both text-only and text-video posts, and a 6\% decline in text-image posts, although the latter remained the most frequent modality overall.

After the invasion, average views per post for Pro-Kremlin channels rose from $15,929$ to $67,442$ (a $323\%$ increase) and for Anti-Kremlin channels from $25,346$ to $104,778$ ($313\%$ increase). Table \ref{tab:modalities_views} presents the distribution of views across the aforementioned multimodal content for both Pro- and Anti-Kremlin channels, highlighting trends before and after the invasion. Notably, anti-Kremlin channels consistently garnered higher viewership than pro-Kremlin channels, especially post-invasion. Further, posts with videos have attracted more views in both channel categories, indicating a strong viewer preference for video content. These trends highlight the dynamic changes in content presentation to increase user engagement, emphasizing the importance of multimodal content.

\subsection{Post Forwarding}
\label{sec:post-forwarding}
Post forwarding on Telegram is a key contributor to user engagement \citep{ng2024exploratory}. Pro-Kremlin channels saw an increase of $313.5\%$ in the average number of post forwarding, rising from $37$ pre-invasion to $153$ post-invasion. Anti-Kremlin channels experienced a $252.4\%$ increase in forwarding, with the average rising from $82$ to $289$ per post. Multimodal posts, which include text, images, and videos, showed more notable increases, from $171$ to $600$ in Pro-Kremlin channels by $250.9\%$, and from $311$ to $1170$ in Anti-Kremlin channels by $276.2\%$. Posts containing videos saw a $231.4\%$ increase, from $140$ to $464$ in Pro-Kremlin channels, and a $283.5\%$ rise, from $231$ to $886$, in Anti-Kremlin channels. Posts in anti-Kremlin channels were forwarded more frequently on average than those in pro-Kremlin channels, highlighting higher user engagement in channels opposing the Kremlin.

\begin{table}[htb!]
    \centering
    \begin{tabular}{p{0.5cm}p{1.4cm}p{1.5cm}p{1.5cm}p{1.4cm}} 
    \toprule[1.5pt]
      \textbf{} & \multicolumn{2}{c}{\textbf{Pro-Kremlin}}  & \multicolumn{2}{c}{\textbf{Anti-Kremlin}}  \\
\cmidrule(r){2-3} \cmidrule(r){4-5}
      \textbf{} & \textbf{Pre- \footnotesize ($\sim$8.4M)} & \textbf{Post- \footnotesize ($\sim$2B)} & \textbf{Pre- \footnotesize ($\sim$5M)} & \textbf{Post- \footnotesize ($\sim$1.2B)} \\
      \midrule[1pt]
      \textbf{\includegraphics[height=1em]{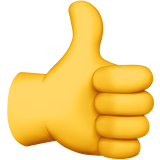}} & 42.39\% & 51.83\% & 39.73\% & 39.12\% \\
      \textbf{\includegraphics[height=1em]{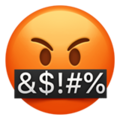}} & 1.6\% & 6.55\% & 3.4\% & 11.14\% \\
      \textbf{\includegraphics[height=1em]{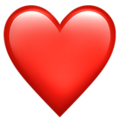}} & 5.05\% & 7.33\% & 7.48\% & 10.11\% \\
      \textbf{\includegraphics[height=1em]{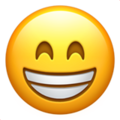}} & 15.14\% & 8.3\% & 9.77\% & 9.0\% \\
      \textbf{\includegraphics[height=1em]{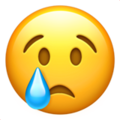}} & 3.78\% & 3.65\% & 3.41\% & 7.09\% \\
      \textbf{\includegraphics[height=1em]{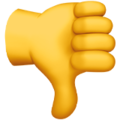}} & 3.85\% & 4.2\% & 10.4\% & 4.52\% \\
      \textbf{\includegraphics[height=1em]{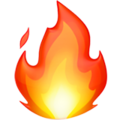}} & 8.86\% & 5.91\% & 5.36\% & 4.1\% \\
      \textbf{\includegraphics[height=1em]{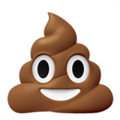}} & 7.88\% & 0.95\% & 9.58\% & 2.77\% \\
      \textbf{\footnotesize others} & 11.45\% & 11.28\% & 10.87\% & 12.15\% \\
      \bottomrule[1.5pt]
    \end{tabular}
    \caption{\emph{Pro-Kremlin} and \emph{Anti-Kremlin} channels most prevalent user reactions in the content for \emph{Pre-} and \emph{Post-invasion}.}
    \label{tab:table-emoji-expanded}
\end{table}

\begin{table*}[htb!]
    \centering
    \begin{tabular}{p{2.5cm}|p{7cm}|p{7cm}}
    \toprule[1.5pt]
    \textbf{N-grams} & \textbf{Before Invasion} & \textbf{After Invasion} \\
    \midrule[1pt]
    \textbf{Unigrams} & 
    Russia, Ukraine, USA, Putin, Moscow, case, military, ruble, elections, political, war, coronavirus, governor, vaccination, court &
    Russia, Ukraine, military, USA, Ukrainian Armed Forces, war, Putin, Moscow, Kyiv, DNR - Donetsk People's Republic, shelling, destroy, Zelensky, sanction, opponent \\
    \midrule
    \textbf{Bigrams} & 
    United Russia, foreign agent, criminal case, Nord Stream, Rostov region, Alexei Navalny, DNR LNR, State Duma elections, Vladimir Zelensky, Sergey Sobyanin, law enforcement agency, price increase, Mikhail Mishustin, Minsk agreement, Alexander Lukashenko &
    Vladimir Putin, combat action, military operation, Kherson region, Kharkov region, White House, Wagner PMC, Ukrainian nationalist, result of shelling, Zelensky, liberate territory, DNR LNR, criminal case, nuclear weapon, Russian oil \\
    \midrule
    \textbf{Trigrams} & 
    performing the function of a foreign agent, mass information, initiate criminal case, Great Patriotic War, China Russia, armed forces of Ukraine, Anthony Blinken, movement forty forty, Russia NATO, detecting new cases of coronavirus &
    special military operation, war in Ukraine, territory of Russia, region launching caliber shells, performing the function of a foreign agent, citizen of Ukraine, reactive system of volley fire, horde of native evil, Great Patriotic War, initiating a criminal case \\
    \bottomrule[1.5pt]
    \end{tabular}%
    \caption{Pro-Kremlin channels most prevalent n-grams in the content of before and after invasion.}
    \label{tab:pro_kremlin_ngrams}
\end{table*}

\subsection{Reactions and Replies}
\label{sec:reactions-and-replies}
Telegram channels can opt to enable or disable user reactions and replies, which impacts public engagement. During our data collection process, distinguishing between posts without reactions due to low engagement and those where reactions were disabled posed a challenge. However, data shows a notable increase in the average number of reactions per post: from $7$ pre-invasion to $991$ post-invasion in Pro-Kremlin channels; from $15$ to $1,533$ in anti-Kremlin channels. Table \ref{tab:table-emoji-expanded} compares the top $10$ emoji used by users pre- and post-invasion for both Pro-Kremlin and Anti-Kremlin channels. In Pro-Kremlin channels, the “\includegraphics[height=1em]{emojis/thumbs-up.png}” emoji dominated over $50\%$ of reactions, while a more diverse range of emoji use was observed in anti-Kremlin channels, including emoji representing strong disagreement, such as “\includegraphics[height=1em]{emojis/na.png}” \citep{kursuncu2019predictive,wijeratne2016emojinet}, with larger volume in Anti-Kremlin channels, post-invasion. The sad emoji “\includegraphics[height=1em]{emojis/crying-face.png}” decreased slightly in Pro-Kremlin channels from $3.78\%$ to $3.65\%$, while an increase was observed in Anti-Kremlin channels from $3.41\%$ to $7.09\%$, indicating increasing emotional response post-invasion.

Before the invasion, both Pro-Kremlin and Anti-Kremlin channels averaged two replies per post, which rose to nine replies per post, indicating higher user engagement through replies post-invasion. In addition, we acknowledge the potential involvement of automated bots in replies or reactions to boost a post's popularity artificially.


\subsection{Content Analysis}
\label{sec:basic-content-analysis}

Our dataset shows specific qualitative differences between Pro-Kremlin and Anti-Kremlin channels, especially pre- and post-invasion. Anti-Kremlin channels typically focus on war atrocities, Russian troop movements, and humanitarian issues. In contrast, Pro-Kremlin channels often try to justify the \textit{“special military operation”} \citep{hanley2023special}. To analyze the basic thematic characteristics of the data, we conducted an n-gram analysis of the text content from these channels both before and after the invasion. We extracted n-grams ($n=1,2,3$) and performed text preprocessing, such as removing punctuation, hyperlinks, numbers, special characters, and Russian stopwords. Additionally, given the extensive use of inflections in Russian, we used lemmatization to standardize different forms of the same word, employing the MyStem package, a Python wrapper for the morphological analysis of the Russian language.

Our analysis of n-grams from both Pro-Kremlin (Table \ref{tab:pro_kremlin_ngrams}) and Anti-Kremlin (Table \ref{tab:anti_kremlin_ngrams}) channels reveals both common and unique phrases potentially distinct narratives. Common phrases include countries, such as \textit{“Russia”}, \textit{“Ukraine”}, \textit{“USA”}; regional and military terms, such as \textit{“Kherson Region”}, \textit{“Kharkov region”}, \textit{“Military”}, \textit{“shelling”}; and leader names such as \textit{“Vladimir Putin”}, \textit{“Vladimir Zelensky”}, \textit{“Joe Biden”}. On the other hand, unique to anti-Kremlin channels are phrases including \textit{“Foreign Agent”}, \textit{“War Crime”}, \textit{“air raid alert”}, \textit{“please keep an eye”}, \textit{“Help”}, which potentially reflect critical perspectives on the conflict and requests for aid. In contrast, pro-Kremlin communications often use phrases unique to these channels, including \textit{“liberate territory”} and \textit{“special military operation”}, indicating a justification of military actions. Further, the term \textit{“special military operation”} was used more frequently than \textit{“war”}, and the phrase \textit{“Great Patriotic War”} also appeared in Pro-Kremlin tri-grams, suggesting a historical framing of current events. This data potentially contains cues for contrasts in how each side portrays the conflict and mobilizes support through online content creation and dissemination. 

\begin{table*}[hbt!]
    \centering
    \begin{tabular}{p{2.5cm}|p{7cm}|p{7cm}}
    \toprule[1.5pt]
    \textbf{N-grams} & \textbf{Before Invasion} & \textbf{After Invasion} \\
    \midrule[1pt]
    \textbf{Unigrams} & 
    Russia, Ukraine, Putin, Navalny, court, Moscow, USA, ruble, detain, police, Belarus, elections, law, criminal, coronavirus &
    Russia, Ukraine, war, military, region, USA, court, strike, missile, sanction, ruble, Kiev, resident, Vladimir, occupant, help, to perish \\
    \midrule
    \textbf{Bigrams} & 
    Foreign Agent, Alexey Navalny, Vladimir Putin, Criminal Case, State Duma Deputy, Russian Authority, Initiate Criminal, Vladimir Zelensky, White House, Supreme Court, State Duma Elections, House Arrest, DNR LNR, Alexander Lukashenko, Recognize Guilty, Ramzan Kadyrov &
    Foreign Agent, Ukraine War, Vladimir Putin, Vladimir Zelensky, Criminal Case, Missile Strike, Kherson Region, Kharkiv Region, War Crime, Joe Biden, Air Raid Alert, Political Circus, Nuclear Weapon, Russian Oil, Military Aid \\
    \midrule
    \textbf{Trigrams} & 
    foreign mass media, initiate a criminal case, Russia Ukraine, Criminal Code of the Russian Federation, cases of COVID actively in the region, United Russia Party, European Court of Human Rights, support Navalny, Anthony Blinken, during the last day in Ukraine, Great Patriotic War &
    Alexey Vladimirovich concerning activities, Russian invasion of Ukraine, special military operation, Donetsk region siren, civil defense shelter, attention air raid sirens, reactive system of volley fire, military aid to Ukraine, Donetsk region all-clear, occupied territory of Ukraine \\
    \bottomrule[1.5pt]
    \end{tabular}%
    \caption{Anti-Kremlin channels most prevalent n-grams in the content of before and after invasion.}
    \label{tab:anti_kremlin_ngrams}
\end{table*}

\section{Data Quality and Qualitative Differences}

The annotation comparison between a native Russian-speaking coder and a non-Russian-speaking coder yielded Cohen's kappa $\kappa$ score of $0.97$, suggesting strong inter-rater agreement. Further, the assessment between the expert and the Russian-speaking and non-Russian-speaking coders revealed kappa $\kappa$ scores of $0.87$ and $0.89$, respectively, indicating substantial agreement with the expert's annotations. These results confirm substantial agreement with the expert's annotations, validating the reliability of our channel classification process.

Our dataset adheres to the \emph{FAIR Data Principles}\footnote{\texttt{https://force11.org/info/the-fair-data- principles/}}. Each data point has a unique identifier accompanied by detailed metadata to enhance findability. The data is stored in a secure repository, which is openly accessible with clearly stated access guidelines. Standardized formats ensure that our dataset is interoperable. Comprehensive documentation provides detailed insights into the data’s collection, processing, and context, supporting its reusability. 

\section{Usage Notes}

The dataset can be accessed via the following Figshare repository\footnote{\texttt{https://doi.org/10.6084/m9.figshare.28785449}}. 
The repository has files in CSV format. This dataset was constructed using public Telegram channels and does not involve any direct interaction with individuals or collection of personally identifiable private data. We have adhered to standard anonymization practices during the data collection and processing stages, ensuring that any personally identifiable information (PII), including user IDs, has been removed. This approach aligns with ethical guidelines to protect the privacy of individuals whose data may be part of the public channels used. It is important to note that this dataset does not include actual images and videos; instead, it contains placeholders indicating the presence of such media. The primary reason for this limitation is the practical challenge of collecting large video files, as Telegram allows users to send video files up to 2 GB in size each. Consequently, this dataset focuses on textual data and metadata, which may limit certain types of multimedia analysis but facilitates more accessible and scalable data handling and processing.

\section{Discussion \& Research Opportunities}
This dataset provides a unique vantage point for analyzing the dynamics of online discourse during the Ukraine-Russia conflict, opening various avenues for answering research questions. Its temporal depth and breadth, which spans over a year before and after the invasion, presents researchers a foundation to explore longitudinal changes in online discourse. By differentiating between pro-Kremlin and anti-Kremlin Telegram channels, the dataset enables comparative analyses of divergent propaganda and counter-narrative strategies, fostering questions into how these narratives shape public perception, political behavior, and transnational information flows.

One major opportunity could be a systematic analysis of narrative formation and transformation in response to real-world geopolitical events. This dataset, enriched with detailed metadata (e.g., timestamps, forwarding activity, emoji reactions), allows researchers to examine how information propagates and evolves across Telegram communities. Contextual content analysis coupling message-level data with known political developments, can reveal how rhetorical strategies respond to, reinforce, or challenge dominant media frames and state-sponsored messaging. This can help scholars investigate phenomena such as agenda-setting, frame alignment, and narrative breach \cite{colleoni2014echo,entman2008theorizing,bawa2024adaptive}.

Further, the dataset can support the application of advanced natural language processing techniques, including topic modeling, stance detection, and sentiment analysis, to identify emergent themes and shifts in thematic patterns \cite{aldayel2021stance,kursuncu2019predictive,kim2024robust,shu2022detecting}. These tools can help uncover the subtle linguistic and semantic patterns embedded in propaganda or dissenting discourse \cite{rashkin2017truth}. When supplemented with entity recognition and relationship extraction, the resulting insights can be integrated with knowledge graphs that structurally map the key actors, themes, and their interrelations across time \cite{hogan2021knowledge}. These knowledge-driven approaches can deepen our understanding of coordinated information operations and opposition narratives by embedding them in a semantically enriched representations \cite{padhi2024enhancing,garg2024just}.

The potential applications of this dataset extend beyond academic research. Journalists, civil society organizations, and policy analysts may leverage the data to track disinformation campaigns, monitor grassroots mobilization, or evaluate the effectiveness of online censorship. Additionally, by studying the temporal interplay between online discourse and offline events, such as military escalations, international diplomacy, or protests—researchers, can better understand how online platforms serve as mirrors or amplifiers of real-world tensions.

Lastly, given Telegram's unique role as a semi-private yet mass-broadcast platform with limited algorithmic moderation, this dataset facilitates examination of how political actors and communities navigate hybrid media ecologies. It invites further inquiry into the affordances and constraints of encrypted platforms in authoritarian and semi-authoritarian settings \cite{gorwa2020unpacking}, and raises important ethical considerations around the governance and transparency of digital public spheres.

\section{Ethical Considerations}
In developing this dataset, we have carefully considered potential negative societal impacts. We recognize that the study of politically sensitive content, especially in the context of conflicts such as the Russia-Ukraine situation, requires a rigorous ethical framework. Hence, we anonymized any personally identifiable information, adhering to ethical standards. We emphasize that any interpretations and conclusions that may be drawn from the dataset are aimed at understanding online communication patterns for academic purposes and should not be used to exacerbate any political conflicts. We are committed to monitoring and addressing any potential misuse that may arise from disseminating this work, ensuring that the research contributes positively to the research community and beyond.

\section{Acknowledgments}

Author Nitin Agarwal acknowledges the support from the U.S. National Science Foundation (OIA-1946391, OIA-1920920), U.S. Army Research Office (W911NF-23-1-0011, W911NF-24-1-0078), U.S. Office of Naval Research (N00014-21-1-2121, N00014-21-1-2765, N00014-22-1-2318), U.S. Air Force Office of Scientific Research (FA9550-22-1-0332), U.S. Air Force Research Laboratory, U.S. Defense Advanced Research Projects Agency, Arkansas Research Alliance, the Jerry L. Maulden/Entergy Endowment at the University of Arkansas at Little Rock, and the Australian Department of Defense Strategic Policy Grants Program.


\bibliography{references}

\begin{thebibliography}{53}
\providecommand{\natexlab}[1]{#1}

\bibitem[{AlDayel and Magdy(2021)}]{aldayel2021stance}
AlDayel, A.; and Magdy, W. 2021.
\newblock Stance detection on social media: State of the art and trends.
\newblock \emph{Information Processing \& Management}, 58(4): 102597.

\bibitem[{Alieva, Kloo, and Carley(2024)}]{alieva2024analyzing}
Alieva, I.; Kloo, I.; and Carley, K.~M. 2024.
\newblock Analyzing Russia’s propaganda tactics on Twitter using mixed methods network analysis and natural language processing: a case study of the 2022 invasion of Ukraine.
\newblock \emph{EPJ Data Science}, 13(1): 42.

\bibitem[{Aro(2016)}]{Aro2016}
Aro, J. 2016.
\newblock The cyberspace war: propaganda and trolling as warfare tools.
\newblock \emph{European View}, 15: 121--132.

\bibitem[{Baumgartner et~al.(2020)Baumgartner, Zannettou, Squire, and Blackburn}]{baumgartner2020pushshift}
Baumgartner, J.; Zannettou, S.; Squire, M.; and Blackburn, J. 2020.
\newblock The pushshift telegram dataset.
\newblock In \emph{Proceedings of the international AAAI conference on web and social media}, volume~14, 840--847.

\bibitem[{Bawa et~al.(2024)Bawa, Kursuncu, Achilov, and Shalin}]{bawa2024adaptive}
Bawa, A.; Kursuncu, U.; Achilov, D.; and Shalin, V.~L. 2024.
\newblock the adaptive strategies of anti-kremlin digital dissent in telegram during the Russian invasion of Ukraine.
\newblock \emph{arXiv preprint arXiv:2408.07135}.

\bibitem[{Center(2023)}]{Levada2023}
Center, L. 2023.
\newblock Main sources of information and popular journalists.
\newblock Accessed: 2024-12-28.

\bibitem[{Cohen(1960)}]{Cohen1960}
Cohen, J. 1960.
\newblock A coefficient of agreement for nominal scales.
\newblock \emph{Educational and Psychological Measurement}, 20: 37--46.

\bibitem[{Colleoni, Rozza, and Arvidsson(2014)}]{colleoni2014echo}
Colleoni, E.; Rozza, A.; and Arvidsson, A. 2014.
\newblock Echo chamber or public sphere? Predicting political orientation and measuring political homophily in Twitter using big data.
\newblock \emph{Journal of communication}, 64(2): 317--332.

\bibitem[{Entman(2008)}]{entman2008theorizing}
Entman, R.~M. 2008.
\newblock Theorizing mediated public diplomacy: The US case.
\newblock \emph{The International Journal of Press/Politics}, 13(2): 87--102.

\bibitem[{for Strategic~Dialogue(2022)}]{ISD2022}
for Strategic~Dialogue, I. 2022.
\newblock A false picture for many audiences: How Russian-language pro-Kremlin Telegram channels spread propaganda and disinformation about refugees from Ukraine.
\newblock Accessed: 2024-12-28.

\bibitem[{Garg et~al.(2024)Garg, Padhi, Jain, Kursuncu, and Kumaraguru}]{garg2024just}
Garg, R.; Padhi, T.; Jain, H.; Kursuncu, U.; and Kumaraguru, P. 2024.
\newblock Just KIDDIN: Knowledge Infusion and Distillation for Detection of INdecent Memes.
\newblock \emph{arXiv preprint arXiv:2411.12174}.

\bibitem[{Ghasiya and Sasahara(2023)}]{ghasiya2023messaging}
Ghasiya, P.; and Sasahara, K. 2023.
\newblock Messaging Strategies of Ukraine and Russia on Telegram during the 2022 Russian invasion of Ukraine.
\newblock \emph{First Monday}.

\bibitem[{Glazunova and Amadoru(2023)}]{Glazunova2023}
Glazunova, S.; and Amadoru, M. 2023.
\newblock "Anti-Regime Influentials" Across Platforms: A Case Study of the Free Navalny Protests in Russia.
\newblock \emph{Media and Communication}, 11: 187--202.

\bibitem[{Gorwa and Guilbeault(2020)}]{gorwa2020unpacking}
Gorwa, R.; and Guilbeault, D. 2020.
\newblock Unpacking the social media bot: A typology to guide research and policy.
\newblock \emph{Policy \& Internet}, 12(2): 225--248.

\bibitem[{Gruzd et~al.(2024)Gruzd, Li, Mai, Ribeiro, and Lab}]{gruzd2024shaping}
Gruzd, A.; Li, Y.; Mai, P.; Ribeiro, T.; and Lab, S.~M. 2024.
\newblock {[Dataset] Shaping the Narratives of the Russia-Ukraine War for Western Audiences: An Exploration of English-language Telegram Channels}.
\newblock Available via Figshare.

\bibitem[{Gunitsky(2015)}]{Gunitsky2015}
Gunitsky, S. 2015.
\newblock Corrupting the cyber-commons: Social media as a tool of autocratic stability.
\newblock \emph{Perspectives on Politics}, 13: 42--54.

\bibitem[{Hanley and Durumeric(2024)}]{hanley2024partial}
Hanley, H.~W.; and Durumeric, Z. 2024.
\newblock Partial mobilization: Tracking multilingual information flows amongst russian media outlets and telegram.
\newblock In \emph{Proceedings of the International AAAI Conference on Web and Social Media}, volume~18, 528--541.

\bibitem[{Hanley, Kumar, and Durumeric(2023{\natexlab{a}})}]{hanley2023special}
Hanley, H.~W.; Kumar, D.; and Durumeric, Z. 2023{\natexlab{a}}.
\newblock " A Special Operation": A Quantitative Approach to Dissecting and Comparing Different Media Ecosystems’ Coverage of the Russo-Ukrainian War.
\newblock In \emph{Proceedings of the International AAAI Conference on Web and social media}, volume~17, 339--350.

\bibitem[{Hanley, Kumar, and Durumeric(2023{\natexlab{b}})}]{hanley2023happenstance}
Hanley, H.~W.; Kumar, D.; and Durumeric, Z. 2023{\natexlab{b}}.
\newblock Happenstance: utilizing semantic search to track Russian state media narratives about the Russo-Ukrainian war on Reddit.
\newblock In \emph{Proceedings of the international AAAI conference on web and social media}, volume~17, 327--338.

\bibitem[{Herasimenka(2022)}]{herasimenka2022movement}
Herasimenka, A. 2022.
\newblock Movement leadership and messaging platforms in preemptive repressive settings: Telegram and the Navalny Movement in Russia.
\newblock \emph{Social Media+ Society}, 8(3): 20563051221123038.

\bibitem[{Hogan et~al.(2021)Hogan, Blomqvist, Cochez, d’Amato, Melo, Gutierrez, Kirrane, Gayo, Navigli, Neumaier et~al.}]{hogan2021knowledge}
Hogan, A.; Blomqvist, E.; Cochez, M.; d’Amato, C.; Melo, G.~D.; Gutierrez, C.; Kirrane, S.; Gayo, J. E.~L.; Navigli, R.; Neumaier, S.; et~al. 2021.
\newblock Knowledge graphs.
\newblock \emph{ACM Computing Surveys (Csur)}, 54(4): 1--37.

\bibitem[{Jurcevic(2019)}]{Jurcevic2019}
Jurcevic, K. 2019.
\newblock Social media-The only voice for oppositional media in Russia?

\bibitem[{Kalsnes and Larsson(2018)}]{Kalsnes2018}
Kalsnes, B.; and Larsson, A.~O. 2018.
\newblock Understanding news sharing across social media: Detailing distribution on Facebook and Twitter.
\newblock \emph{Journalism Studies}, 19: 1669--1688.

\bibitem[{Karalis(2024)}]{Karalis2024}
Karalis, M. 2024.
\newblock Russia-Ukraine through the eyes of social media.
\newblock Accessed: 2024-12-28.

\bibitem[{Khaund et~al.(2020)Khaund, Hussain, Shaik, and Agarwal}]{Khaund2020}
Khaund, T.; Hussain, M.~N.; Shaik, M.; and Agarwal, N. 2020.
\newblock Telegram: Data collection, opportunities and challenges.
\newblock In \emph{Proceedings of the Annual International Conference on Information Management and Big Data}, 513--526. Springer International Publishing.

\bibitem[{Kim et~al.(2024)Kim, Mosallanezhad, Cheng, Mancenido, and Liu}]{kim2024robust}
Kim, N.; Mosallanezhad, D.; Cheng, L.; Mancenido, M.~V.; and Liu, H. 2024.
\newblock Robust Stance Detection: Understanding Public Perceptions in Social Media.
\newblock In \emph{International Conference on Advances in Social Networks Analysis and Mining}, 21--37. Springer.

\bibitem[{Kloo, Cruickshank, and Carley(2024)}]{kloo2024cross}
Kloo, I.; Cruickshank, I.~J.; and Carley, K.~M. 2024.
\newblock A Cross-Platform Topic Analysis of the Nazi Narrative on Twitter and Telegram During the 2022 Russian Invasion of Ukraine.
\newblock In \emph{Proceedings of the International AAAI Conference on Web and Social Media}, volume~18, 839--850.

\bibitem[{Krever and Chernova(2023)}]{krever2023wagner}
Krever, M.; and Chernova, A. 2023.
\newblock Wagner chief admits to founding Russian troll farm sanctioned for meddling in US elections.
\newblock Accessed: 2025-04-13.

\bibitem[{Kursuncu et~al.(2019)Kursuncu, Gaur, Lokala, Thirunarayan, Sheth, and Arpinar}]{kursuncu2019predictive}
Kursuncu, U.; Gaur, M.; Lokala, U.; Thirunarayan, K.; Sheth, A.; and Arpinar, I.~B. 2019.
\newblock Predictive analysis on Twitter: Techniques and applications.
\newblock In \emph{Emerging research challenges and opportunities in computational social network analysis and mining}, 67--104. Springer Nature.

\bibitem[{Kursuncu et~al.(2021)Kursuncu, Purohit, Agarwal, and Sheth}]{Kursuncu2021}
Kursuncu, U.; Purohit, H.; Agarwal, N.; and Sheth, A. 2021.
\newblock When the bad is good and the good is bad: Understanding cyber social health through misc behavioral change.
\newblock \emph{IEEE Internet Computing}, 25(1): 6--11.

\bibitem[{Lai, Toriumi, and Yoshida(2024)}]{lai2024multilingual}
Lai, C.; Toriumi, F.; and Yoshida, M. 2024.
\newblock A multilingual analysis of pro Russian misinformation on Twitter during the Russian invasion of Ukraine.
\newblock \emph{Scientific Reports}, 14(1): 10155.

\bibitem[{Linvill and Warren(2020)}]{linvill2020troll}
Linvill, D.~L.; and Warren, P.~L. 2020.
\newblock Troll factories: Manufacturing specialized disinformation on Twitter.
\newblock \emph{Political Communication}, 37(4): 447--467.

\bibitem[{Murauskaite(2023)}]{Murauskaite2023}
Murauskaite, E. 2023.
\newblock \emph{U.S. Arms Transfers to Ukraine: Impact Assessment}.
\newblock College Park, MD: START.

\bibitem[{Ng et~al.(2024)Ng, Kloo, Clark, and Carley}]{ng2024exploratory}
Ng, L. H.~X.; Kloo, I.; Clark, S.; and Carley, K.~M. 2024.
\newblock An exploratory analysis of COVID bot vs human disinformation dissemination stemming from the Disinformation Dozen on Telegram.
\newblock \emph{Journal of Computational Social Science}, 1--26.

\bibitem[{Padhi et~al.(2024)Padhi, Kursuncu, Kumar, Shalin, and Fronczek}]{padhi2024enhancing}
Padhi, T.; Kursuncu, U.; Kumar, Y.; Shalin, V.~L.; and Fronczek, L.~P. 2024.
\newblock Enhancing Cross-Modal Contextual Congruence for Crowdfunding Success using Knowledge-infused Learning.
\newblock In \emph{2024 IEEE International Conference on Big Data (BigData)}, 1844--1853. IEEE.

\bibitem[{Park et~al.(2022)Park, Mendelsohn, Field, and Tsvetkov}]{Park2022}
Park, C.~Y.; Mendelsohn, J.; Field, A.; and Tsvetkov, Y. 2022.
\newblock Challenges and opportunities in information manipulation detection: An examination of wartime Russian media.
\newblock In \emph{Findings of the Association for Computational Linguistics (EMNLP)}, 5209--5235.

\bibitem[{Pierri et~al.(2023)Pierri, Luceri, Jindal, and Ferrara}]{pierri2023propaganda}
Pierri, F.; Luceri, L.; Jindal, N.; and Ferrara, E. 2023.
\newblock Propaganda and misinformation on Facebook and Twitter during the Russian invasion of Ukraine.
\newblock In \emph{Proceedings of the 15th ACM web science conference 2023}, 65--74.

\bibitem[{Polyanskaya, Krivov, and Lomko(2003)}]{Polyanskaya2003}
Polyanskaya, A.; Krivov, A.; and Lomko, I. 2003.
\newblock Virtualnoye Oko Stareshevo Brata [The Virtual Eye of Big Brother].
\newblock \emph{Vestnik misc}, 9: 320.

\bibitem[{Ramani(2023)}]{Ramani2023}
Ramani, S. 2023.
\newblock \emph{Putin's War on Ukraine: Russia’s Campaign for Global Counter-Revolution}.
\newblock Hurst Publishers.

\bibitem[{Rashkin et~al.(2017)Rashkin, Choi, Jang, Volkova, and Choi}]{rashkin2017truth}
Rashkin, H.; Choi, E.; Jang, J.~Y.; Volkova, S.; and Choi, Y. 2017.
\newblock Truth of varying shades: Analyzing language in fake news and political fact-checking.
\newblock In \emph{Proceedings of the 2017 conference on empirical methods in natural language processing}, 2931--2937.

\bibitem[{Re:Russia(2024)}]{ReRussia2024}
Re:Russia. 2024.
\newblock What does the Kremlin fear? An analysis of protest potential in Russian society.
\newblock Accessed: 2024-12-28.

\bibitem[{Sherstoboeva(2024)}]{Sherstoboeva2024}
Sherstoboeva, E. 2024.
\newblock Russian Bans on ‘Fake News’ about the war in Ukraine: Conditional truth and unconditional loyalty.
\newblock \emph{International Communication Gazette}, 86: 36--54.

\bibitem[{Shu and Liu(2022)}]{shu2022detecting}
Shu, K.; and Liu, H. 2022.
\newblock \emph{Detecting fake news on social media}.
\newblock Springer Nature.

\bibitem[{Smyth and Oates(2017)}]{Smyth2017}
Smyth, R.; and Oates, S. 2017.
\newblock Mind the gaps: Media use and mass action in Russia.
\newblock In \emph{State Against Civil Society}, chapter~4, 127--147. Routledge.

\bibitem[{Spaiser et~al.(2017)Spaiser, Chadefaux, Donnay, Russmann, and Helbing}]{Spaiser2017}
Spaiser, V.; Chadefaux, T.; Donnay, K.; Russmann, F.; and Helbing, D. 2017.
\newblock Communication power struggles on social media: A case study of the 2011–12 Russian protests.
\newblock \emph{Journal of Information Technology and Politics}, 14: 132--153.

\bibitem[{Starbird, Arif, and Wilson(2019)}]{starbird2019disinformation}
Starbird, K.; Arif, A.; and Wilson, T. 2019.
\newblock Disinformation as collaborative work: Surfacing the participatory nature of strategic information operations.
\newblock \emph{Proceedings of the ACM on human-computer interaction}, 3(CSCW): 1--26.

\bibitem[{Survey(2022)}]{OPORA2022}
Survey, O. 2022.
\newblock Media Consumption of Ukrainians in a Full-Scale War.
\newblock Accessed: 2024-12-28.

\bibitem[{Telethon(2024)}]{Telethon2024}
Telethon. 2024.
\newblock \emph{Telethon’s Documentation}.
\newblock Accessed: 2024-12-28.

\bibitem[{TGStat(2024)}]{TGStat2024}
TGStat. 2024.
\newblock Telegram channels and groups catalog.
\newblock Accessed: 2024-12-28.

\bibitem[{Times(2023)}]{MoscowTimes2023}
Times, T.~M. 2023.
\newblock Telegram surpasses WhatsApp traffic volume in Russia.
\newblock Accessed: 2024-12-28.

\bibitem[{{UK Government}(2022)}]{ukgov2022trollfactory}
{UK Government}. 2022.
\newblock UK exposes sick Russian troll factory plaguing social media with Kremlin propaganda.
\newblock Accessed: 2025-04-13.

\bibitem[{Voorveld et~al.(2018)Voorveld, Van~Noort, Muntinga, and Bronner}]{Voorveld2018}
Voorveld, H. A.~M.; Van~Noort, G.; Muntinga, D.~G.; and Bronner, F. 2018.
\newblock Engagement with social media and social media advertising: The differentiating role of platform type.
\newblock \emph{Journal of Advertising}, 47: 38--54.

\bibitem[{Wijeratne et~al.(2016)Wijeratne, Balasuriya, Sheth, and Doran}]{wijeratne2016emojinet}
Wijeratne, S.; Balasuriya, L.; Sheth, A.; and Doran, D. 2016.
\newblock Emojinet: Building a machine readable sense inventory for emoji.
\newblock In \emph{Social Informatics: 8th International Conference, SocInfo 2016, Bellevue, WA, USA, November 11-14, 2016, Proceedings, Part I 8}, 527--541. Springer.

\end{thebibliography}




\end{document}